# Why are differential equations used for expressing the laws of physics?

## Shabnam Siddiqui

*Institute of Micromanufacturing and Department of Physics*

*Louisiana Tech University, Ruston, LA USA*

**Summary:** Almost all theories of physics have expressed physical laws by means of differential equations. One can ask: why differential equations? What is special about them? This article addresses these questions and is presented as an inquiry-based lecture, where students and a teacher are engaged in discussion. It has two goals: (1) To help undergraduate students understand the rationale behind the use of differential equations in physics, (2) To show how meaningful and interactive presentation of mathematics can help students take pleasure in learning physics.

A teacher enters a class-room.

Teacher: Let me begin this lecture with a simple question. "Are there any mathematical similarities between the laws of motion, electricity, magnetism, and optics etc?"

There is complete silence in the class.

Teacher: "If you thought that the answer to this question is that most of these laws of Physics are expressed in terms of differential equations then, yes, you are right. But why are differential equations used to express physical laws? I mean, what is special about them?"

Again there is complete silence in the classroom, and the students look surprised.

Teacher: "We are going to understand the reasons behind the use of differential equations for expressing the Laws of Physics.

The Teacher moves toward the blackboard and writes the following question:

> **Why are differential equations used for expressing physical laws?**

After writing the question, the teacher again poses another question.

Teacher: In order to answer this question we need to understand few definitions and their meanings, and poses another question. Can anyone tell me what does a physical law mean?

Student: "A law is a set of rules followed by a system."

Teacher: "Right! Such rules are used to explain a phenomenon exhibited by a system and are usually expressed as a mathematical relation. Such rules are observed by repeated experimentation. For example, Newton's laws are the set of rules followed by a system exhibiting motion. The second law of motion is expressed in the form of second-order ordinary differential equation." The teacher writes the following equation on the board.

$$F(x) = m\frac{d^2x}{dt^2}$$ Eqn. [1]

Teacher: "In this equation, 'x' is the distance of a point mass at any instant 't' from a point O taken as the origin on a straight line. An equation of this kind is called a differential equation because this relationship involves a derivative ($\frac{d^2x}{dt^2}$). The order of a differential equation is the same as the order of the highest derivative, thereby Eqn. [1] is a second order differential equation. Here you can ask why force is defined in terms of a second order differential equation, and not first order?

In order to answer this question, let us first try to understand the conditions that mathematical relationships must satisfy in order to express a physical law, and later we will see how these conditions are satisfied by differential equations."

The teacher writes following on the blackboard.

> **Necessary conditions for expressing a physical law**
>
> (1) The mathematical relation must be sufficiently general
> (2) It must define connections between neighboring points
> (3) It must imply the continuity of change.

Teacher: "Let us make an attempt to understand the first necessary condition for expressing a physical law." The teacher points to the blackboard. "Can anyone tell me the physical meaning of this condition and why it is necessary?"

A second student: "Since a physical law is universal, the mathematical relationship expressed by it must be general".

Teacher: "Yes, it should be general so that it is open to a whole class of possibilities, so that by changing the initial conditions of the system the physical law must not change. The set of rules are fixed and system follows these rules under different circumstances. However, the laws can break-down in some situations, and therefore a law is true only for a subset of conditions. For example, all macroscopic systems using Newton's laws of motion one can find the exact position and momentum of an object at any instant of time whereas for quantum sized systems one cannot determine momentum and position values at any instant of time with full certainty. Another example is that of ohms law, for certain systems such as copper wire one can find linear relationship between current and voltage, whereas for systems such as transistors such relationship breakdowns.

Teacher: "Now, let us move on to the necessary conditions number two and three and try to understand their meaning. According to classical physics, all systems change continuously as time passes. A physical law expresses a permanent relationship between the state of the system at the present moment and its state immediately following that moment. Therefore, a law must define the connection between successive space-time points. As we can see in Eqn. [1], Newton's second law of motion defines a relationship that connects the velocity of the body at a present space-time point with the velocity at a point in the immediate neighborhood of the first space-time point. A majority of physical processes change continuously, and therefore a law must express this continuity of change (requirement number three), in a *causal chain*. In a continuous process, the sequence of states assumed by a physical system forms a continuous chain, which means that the change in state of the system over an infinitesimal period of time is itself infinitesimal. The laws controlling such processes must therefore exhibit continuity of change."

Another student: "A law expresses the rules that a system must follow in order to change from state A to state B where state B is the state of the system immediately following state A".

Teacher: "Yes, precisely".

Student: "What about the processes that does not change continuously?"

Teacher: "In fact, there are processes in nature that do not change continuously. But it is one of the most important assumptions in classical physics that all processes in nature change continuously. Even in a collision between two balls, during which the paths of the balls change abruptly, this assumption is maintained. Such abrupt change is considered to be continuous even though very rapid".

Teacher: "Let us now see how these requirements are satisfied by differential equations, ordinary differential equations.

Let us recall the following mathematical relationship:

$y = f(x)$                                                          Eqn. [2]

Here y is a continuous function of a real variable x. Consider the following function

$y = \pm\sqrt{16 - x^2}$                                       Eqn. [3]

This equation represents a circle of radius 4, with center at origin. By differentiating the equation with respect to 'x' we obtain the following relationship:

$y \dfrac{dy}{dx} + x = 0$                                      Eqn.[4]

Now, both equations, Eqn. [3] & Eqn. [4] define a circle. These two equations are not equivalent, Eqn. [3] defines a circle of radius 4, whereas Eqn. [4] defines all circles (not merely a circle of radius 4) having the center of the circle as the origin. We can see this by integrating Eqn. [4], and derive the following relationship.

$y = \pm\sqrt{R^2 - x^2}$                                    Eqn. [5]

This relationship defines all circles having their center as the origin. In mathematical language, all circles regardless of their radius satisfy the differential Eqn. [4]. This important property of defining a class and not a particular member of the class is characteristic of all differential equations. Therefore, we see that requirement number one for a physical law is satisfied."

Teacher: "Can anyone explain how requirements number two and three are satisfied by ordinary differential equations?"

Student: "I think the derivative of a differential equation sets-up a connection between the values of neighboring space-time points".

Teacher: "Absolutely. But, what about the condition number three?"

Student: "The existence of derivatives at every space-time point confirms that the system is changing continuously."

Teacher: "Yes, you are absolutely right. Therefore, a law expressed in terms of a differential equation always satisfy condition number two and three.

Let us now go back to Newton's equation for the force, and try to answer why force is expressed in terms of a second order differential equation. Let us integrate this equation (Eqn. [1]) on both sides with respect to time and see what happens. We get following equation:

$$\int_0^t F(x)dt = m\frac{dx}{dt} + C \qquad \text{Eqn. [6]}$$

The above equation is not a closed form differential equation. In order for the above equation to be a first-order closed form differential equation, the form of the force $F(x)$ needs to be defined. Thereby, Eqn. [6] cannot be reduced to a first-order differential equation for a general force $F(x)$, and condition number one is not satisfied.

Thus, only using a second order differential equation, Newton's second law can be expressed for a general force $F(x)$ as in Eqn. [1].

However, for $F(x)/m=f=$constant, this equation simplifies to the following form:

$$ft = \frac{dx}{dt} + C \qquad \text{Eqn. [7]}$$

and all the typical equations of motion for a constant force can be derived from this equation. Eqn. [7] describes Newton's second law of motion but only for a constant force, and is not sufficiently general.

**Only for a constant force, Newton's second law can be written in the form of first order differential equation.** Teacher repeats.

Teacher stops talking

There is complete silence in the class.

Teacher: "Are you still listening to my argument?"

Class laughs.

Teacher resumes.

"Any questions? Well, it is time to end our discussion.

Here is something to think about and have fun, and we can discuss this in our next class."

Teacher writes a question on the black board.

**Question: Integrate Eqn. [7], and plot this equation and Eqn. [7] for x(t) Vs t, and v(t) Vs t, by choosing your own constants. Look for continuity in the plot, and convince yourself whether other two conditions are satisfied or not.**

Acknowledgment

I would like to thank Ian Wiley (Advanced Diamond Technologies, INC) for improving technical writing and Dr. Lee Sawyer (Louisiana Tech University) for his valuable inputs.